\documentstyle[12pt,thmsa,sw20aip]{article}

\input{tcilatex}
\begin{document}

\title{Gravitomagnetism, clocks and geometry}
\author{A. Tartaglia \\
Dip. Fisica, Politecnico, Torino, Italy\\
e-mail: tartaglia@polito.it}
\maketitle

\begin{abstract}
New techniques to evaluate the clock effect using light are described. These
are based on the flatness of the cylindrical surface containing the world
lines of the rays constrained to move on circular trajectories about a
spinning mass. The effect of the angular momentum of the source is
manifested in the fact that inertial observers must be replaced by local non
rotating observers. Starting from this an exact formula for circular
trajectories is found. Numerical estimates for the Earth environment show
that light would be a better probe than actual clocks to evidence the
angular momentum influence. The advantages of light in connection with some
principle experiments are shortly reviewed.
\end{abstract}

\section{Introduction}

The gravitomagnetic clock effect (synthetically gclock effect) is in a sense
a phenomenon announced since the early days of relativity\cite{einstein}. It
is due to the fact that the proper time measured by a clock is indeed the
four dimensional ''length'' of the world line of the clock between two given
events: different allowed four dimensional paths between the same pair of
events correspond to different proper time intervals and produce
desynchronization of initially synchronous clocks.

When treating rotating systems (and clocks) an additional feature one finds
is that a global splitting between space and time is no more possible: time
on a rotating platform becomes polytropic\cite{rt(poly)} in the sense that
the space hypersurface of a rotating observer is a helicoid and it
intercepts the world line of the observer in an infinite number of events:
these are all nominally ''simultaneous'' though corresponding to different
proper times of the observer. This fact does not really affect the nature
and geometric interpretation of the clock effect. 

The revolving ''clocks'' used to measure proper times may in fact be various
types of particles, such as electrons\cite{electrons}, neutrons\cite
{neutrons}, Cooper pairs\cite{cooper}, calcium atoms\cite{calcium}, which
have been used to test the Sagnac effect for quantum objects. Of course one
can use light as a ''clock'' finding an asymmetry between pro- and
retrograde rotation manifested in a phase difference rather than a
difference in proper times, now identically null. This is the very Sagnac
effect, known since 1913 or even before and named after G. Sagnac\cite
{sagnac}, who interpreted it with an antirelativistic and ''etherial'' (in
the sense of the ether theory) attitude. The Sagnac effect is nothing else
than a specialization of what we are speaking about since the beginning of
this section and earned a whole literature for itself during many decades%
\cite{stedman}; the simplest and most complete description of it is indeed
the geometrical one\cite{anandan}\cite{rt(poly)}. An effect with actual
clocks, and in particular the synchrony defect due to the time polytropy,
has been evidenced for the first time with satellites in 1985\cite
{satellites}.

The introduction of nonspinning or spinning masses in the empty space time
does not in principle modify the situation, but for computational
complications. The latter case (spinning sources of gravity) is however the
one to which the expression ''gravitomagnetic clock effect'' has been
reserved. The name comes from the fact that, in weak field approximation,
the gravitational field may be decomposed into a gravitoelectric (radial)
field and a gravitomagnetic (solenoidal) field in analogy with
electromagnetism: the gclock effect is determined by the gravitomagnetic
component. This description and treatment has been particularly emphasized
by Mashhoon and others\cite{mashhoon1} \cite{mashhoon2}, but a general
treatment is in principle rather simple and can be given before any
approximation\cite{tartagliaGRG}.

The importance of the gclock effect is in the hope that it can be used to
test the influence of the angular momentum of the source on the
gravitational field. In fact the revolution time of a freely orbiting clock
on a circular trajectory around a spinning mass is different according to
whether the orbit is prograde or retrograde. The difference between the two
periods expressed as the times needed to recover a fixed azimuth in the
inertial frame of a distant observer is\cite{mashhoon2}: 
\begin{equation}
\delta t=4\pi \frac{a}{c}  \label{assoluto}
\end{equation}
where $a$ is a length given by the ratio between the angular momentum of the
source $J$ and its mass $M$ multiplied by $c$: $a=\frac{J}{Mc}$. The result (%
\ref{assoluto}) is true for any axisymmetric metric with an $a$ value
independent from time\cite{tartagliaGRG}. The independence of (\ref{assoluto}%
) from the radius of the orbit is a remarkable feature; furthermore the
numerical value of $\delta t$ for the Earth is rather promising being in the
order of $\sim 10^{-7}$ s.

When adopting the viewpoint of an observer in the vicinity of the central
mass the situation worsens; then one should use proper times of the clocks
and refer for instance to specific events, such as the conjunctions of
counterorbiting objects. In that case the time difference on the clocks
around the Earth, $\delta \tau $, is in the order of $\sim 10^{-16}$ s \cite
{mashtart}\cite{tartagliaCQG}.

Besides the actual numbers that should be measured (which, at least in the
case of $\delta t$, are indeed big), a real experiment around the Earth
encounters a series of difficulties not easy to be surmounted\cite{mashhoon2}
\cite{mashtart}, one of them is the stability and coincidence between
opposite orbits. Considering this specific problem electromagnetic waves
come again on stage as probes more interesting than actual clocks. In fact a
single ring of orbiting mirrors could be used to reflect at the same instant
both prograde and retrograde ''light'' beams in a sort of Sagnac experiment
in space. The influence of gravity on the issue of such an experiment can be
worked out as a general relativistic correction of the classical effect\cite
{tartaglia2}, but the problem can be better solved in principle and in
general using the method outlined in ref. \cite{tartagliaGRG}. This is
precisely the aim of this paper; in sect. II the essence of the method is
recalled; in sect. III the specific case of the Sagnac effect is treated;
sect. IV contains the approximated results valid for the Earth environment;
finally in sect. V some conclusions will be drawn.

\section{Geometric vision of the gclock effect}

The method proposed is based on the fact that in any axially symmetric and
stationary space time any constant $r$ and $\theta $ world line is drawn on
the surface of a cylinder which is indeed flat, no matter what the global
curvature is: on that surface simple Minkowskian geometry works. In
particular when the coordinate angular velocity $\omega =d\phi /dt$ is fixed
the world line on the cylinder's surface is a helix, which becomes a
straight line in the development of the cylinder in a plane.

The situation is most simply described graphically: see fig. 1, which shows
the opened cylindrical world tube. The world lines of steadily rotating
objects are helices which in turn become straight lines when opening the
cylinder on a plane. When a line reaches the boundary of the cut at $\pi $
or $-\pi $ it bounces back or forth by $2\pi $ continuing on the other side.

\FRAME{dtbpFU}{3.5976in}{2.7008in}{0pt}{\Qcb{Fig. 1 Scheme of the opened
world tube on whose surface the world lines of the observer and the clocks
lie. Points {\it a} and {\it a' }correspond to each other. {\it Oaa'B}
represents a prograde light ray; {\it Oa'aA }is a retrograde light ray; {\it %
OB} is the world line of the observer. The segment {\it AB} visualizes the
proper time delay which is indeed the Sagnac effect.}}{\Qlb{prima}}{%
prdfig11.gif}{%
}

Considering a pair of objects revolving in opposite directions with the same
coordinate speed, their world lines are symmetric with respect to a static
observer at the origin (vertical straight line there) and cross each other
again at the same event for such observer: the revolution times both in the
objects proper times and in the observer's time are exactly the same for
both. If however the observer is in turn steadily rotating his world line is
an oblique straight line and the intersecting events with the world lines of
the objects do not coincide any more: the revolution times measured by the
observer differ from each other and so do the proper times of the two
objects. The interval between {\it A }and{\it \ B, }in fig. 1{\it , }is then
proportional to the proper time lapse (for the observer) between the
completions of one revolution started simultaneously by the two probes. This
description is valid for light beams too: in this case the $AB$ interval is
an actual measure of the Sagnac effect.

When the central body possesses an angular momentum, the equivalent in its
vicinity of the static inertial observers of the previous case are the so
called Locally Non Rotating Observers (LNRO)\cite{straumann}, i. e. the
observers to whom radially infalling matter appears locally as non rotating.
The LNRO's themselves are indeed rotating as seen by a distant inertial
observer; their coordinate angular speed is: 
\begin{equation}
\Omega =-g_{t\phi }/g_{\phi \phi }  \label{LNRO}
\end{equation}
(the $g$'s are elements of the metric and the use of polar coordinates is
understood). In practice the considerations about the flatness of the
cylindrical surface containing the world lines of constant $r$ and $\theta $
remain unchanged, but now the world lines of co-rotating and
counter-rotating light rays have a different inclination to cope with the
role of an LNRO and his skew world line; in other words after leaving the
LNRO at $O$ the two beams must cross each other again on the (oblique) world
line of the LNRO, in order it to be equivalent to the static observer of
rotation free space times. Graphically the situation is the one shown in
fig. 2.

\FRAME{dtbpFU}{3.544in}{2.6636in}{0pt}{\Qcb{Fig 2. The scheme is the same as
in fig. 1, but now the light rays are represented by different inclinations
on the right and on the left of the origin. Events $A^{\prime }$ and $%
B^{\prime }$ delimitate the proper time span between one complete revolution
of retrograde and prograde light rays. The rays meet each other at the world
line of the LNRO passing through the origin event.}}{\Qlb{seconda}}{%
aipfig2.gif}{
}

Now the entity of the Sagnac effect is measured by the interval between $%
A^{\prime }$ and $B^{\prime }$ which is clearly different from the one
between $A$ and $B$ in fig. 1, depending on the value of the parameter $a$
which is hidden in the $g_{t\phi }$ element of the metric tensor.

\section{Influence of the angular momentum on the Sagnac effect}

What is visually determined by the simple graphic treatment of the previous
section can be put in analytical form by the use of Minkowskian geometry in
two dimensions. Let us restrict our study to light rays and the Sagnac
effect in axisymmetric metrics. Events $A$ and $B$ in fig. 1 ($A^{\prime }$
and $B^{\prime }$ in fig. 2) are localized as the intersections of two
straight lines: the world line of the observer and the world line of a light
beam. When an angular momentum is absent the former's equation is $t=\phi
/\omega _{o}$, where $\omega _{o}$ is the observer's angular coordinate
speed. The equation of the world line of a light ray constrained to move
along a circular trajectory is in turn $t=\left( 2\pi -\phi \right) /\omega
_{l0}$ for the counter-rotating beam and $t=\left( \phi +2\pi \right)
/\omega _{l0}$ for the co-rotating one; the equations reproduce the lines
respectively ascending to the left through $a$ and ascending to the right
through $a^{\prime }$ in fig. 1. The coordinate angular speed of light for a
circular path in a plane orthogonal to the symmetry axis (let us say it is
the equatorial plane of the system) is $\omega _{l0}=\sqrt{-g_{tt}/g_{\phi
\phi }}$; this comes from a null interval at constant $r$ and $\theta $,
i.e. from $0=g_{tt}dt^{2}+g_{\phi \phi }d\phi ^{2}$, posing $d\phi
/dt=\omega _{l0}$. The intersections between the straight world lines of the
observer and the beams identify points $A$ (counter-rotating beam) and $B$
(corotating beam) with their $t$ and $\phi $ coordinates: 
\begin{equation}
\left\{ 
\begin{array}{c}
\phi _{A,B}=2\pi \frac{\omega _{o}}{\omega _{l0}\pm \omega _{o}} \\ 
t_{A,B}=\frac{2\pi }{\omega _{l0}\pm \omega _{o}}
\end{array}
\right.  \label{prima}
\end{equation}
The $+$ ($-$) sign corresponds to event $A$ ($B$). The proper time
corresponding to the interval $OA$ ($OB$) in fig. 1 is in practice the
length (divided by $c$) of the hypotenuse of a right-angled triangle in a
two dimensional flat Minkowski space time, whose other sides are $\sqrt{%
g_{tt}t_{A,B}^{2}}$ and $\sqrt{g_{\phi \phi }\phi _{A,B}^{2}}$. The explicit
result is: 
\[
\tau _{A,B}=\frac{2\pi }{c\left( \omega _{l0}\pm \omega _{o}\right) }\sqrt{%
g_{tt}+g_{\phi \phi }\omega _{o}^{2}} 
\]
and the Sagnac delay turns out to be 
\begin{equation}
\delta \tau =\tau _{B}-\tau _{A}=4\pi \frac{\omega _{o}}{c\left( \omega
_{l0}^{2}-\omega _{o}^{2}\right) }\sqrt{g_{tt}+g_{\phi \phi }\omega _{o}^{2}}%
=4\pi \frac{\omega _{o}}{c}\sqrt{\frac{-g_{\phi \phi }}{\omega
_{l0}^{2}-\omega _{o}^{2}}}  \label{sagnac0}
\end{equation}
The $g_{tt}$ term has been eliminated thanks to the explicit expression of $%
\omega _{l0}$.

When $g_{t\phi }\neq 0$ the coordinate angular speed of light on a circular
path is obtained from $0=g_{tt}+2g_{t\phi }\omega +g_{\phi \phi }\omega ^{2}$%
. There are two solutions: 
\begin{equation}
\omega _{l\pm }=\sqrt{\Omega ^{2}+\omega _{l0}^{2}}\pm \Omega  \label{luce}
\end{equation}
($+$ is for prograde motion, $-$ is for retrograde motion).

Now the intersection method leads to: 
\begin{equation}
\left\{ 
\begin{array}{c}
\phi _{A^{\prime },B^{\prime }}=\frac{2\pi \omega _{o}}{\omega _{l\mp }\pm
\omega _{o}} \\ 
t_{A^{\prime },B^{\prime }}=\frac{2\pi }{\omega _{l\mp }\pm \omega _{o}}
\end{array}
\right.  \label{inmoto}
\end{equation}

The proper observer's intervals of time are obtained as before, but for the
fact that now we have to use the equivalent of Carnot's theorem of Euclidean
geometry rather than the equivalent of Pythagora's: 
\[
\tau _{A^{\prime },B^{\prime }}=\frac{2\pi }{c\left( \omega _{l\mp }\pm
\omega _{o}\right) }\sqrt{g_{tt}+2g_{t\phi }\omega _{o}+g_{\phi \phi }\omega
_{o}^{2}} 
\]

Finally the Sagnac delay is: 
\begin{eqnarray}
\delta \tau &=&\tau _{B^{\prime }}-\tau _{A^{\prime }}=\frac{4\pi }{c}\frac{%
\omega _{o}-\Omega }{\omega _{l0}^{2}+2\Omega \omega _{o}-\omega _{o}^{2}}%
\sqrt{g_{tt}+2g_{t\phi }\omega _{o}+g_{\phi \phi }\omega _{o}^{2}}
\label{deltatau} \\
&=&\frac{4\pi }{c}\frac{-g_{\phi \phi }\omega _{o}-g_{t\phi }}{\sqrt{%
g_{tt}+2g_{t\phi }\omega _{o}+g_{\phi \phi }\omega _{o}^{2}}}  \nonumber
\end{eqnarray}
The parameter $\Omega $ has been eliminated using the explicit expression
for the coordinate angular speed of an LNRO.

If now we pose $\omega _{o}=\Omega $, we consistently verify that it is: 
\[
\delta \tau _{LNRO}=0 
\]
A truly inertial observer ($\omega _{o}=0$) would find: 
\begin{equation}
\delta t=-\frac{4\pi }{c}\frac{g_{t\phi }}{\sqrt{g_{tt}}}  \label{ossin}
\end{equation}

Formula (\ref{deltatau}) treats $\omega _{o}$ and $r$ as independent
parameters; the situation changes when the observer is freely orbiting. For
a circular geodesic trajectory in the equatorial plane there are two
solutions for $\omega _{o}$ \cite{tartagliaGRG} (again $+$ is pro- and $-$
is retrograde motion of the observer): 
\begin{equation}
\omega _{o\pm }=\frac{-g_{t\phi ,r}\pm \sqrt{g_{t\phi ,r}^{2}-g_{\phi \phi
,r}g_{tt,r}}}{g_{\phi \phi ,r}}  \label{orbita}
\end{equation}
Commas denote ordinary partial differentiation.. Correspondingly (\ref
{deltatau}) becomes: 
\begin{equation}
\delta \tau _{\pm }=\frac{4\pi }{c}\frac{-g_{\phi \phi }\omega _{o\pm
}-g_{t\phi }}{\sqrt{g_{tt}+2g_{t\phi }\omega _{o\pm }+g_{\phi \phi }\omega
_{o\pm }^{2}}}  \label{completo}
\end{equation}

\section{Terrestrial environment}

Considering the situation in the surroundings of our planet the appropriate
metric is the one corresponding to a weak axisymmetric field. Actually this
means that: 
\begin{eqnarray*}
g_{tt} &=&c^{2}\left( 1-2G\frac{M}{c^{2}r}\right) \\
g_{t\phi } &=&2aG\frac{M}{cr} \\
g_{\phi \phi } &\simeq &-r^{2}
\end{eqnarray*}

Consequently it is: 
\begin{eqnarray*}
\Omega &=&2aG\frac{M}{cr^{3}} \\
\omega _{l0} &=&\frac{c}{r}\sqrt{1-2G\frac{M}{c^{2}r}}
\end{eqnarray*}

Now using (\ref{deltatau}) we obtain:

\begin{equation}
\delta \tau \simeq \frac{4\pi }{c^{2}}\frac{r^{2}\omega _{o}-2aG\frac{M}{cr}%
}{\sqrt{1-2G\frac{M}{c^{2}r}+4aG\frac{M}{c^{3}r}\omega _{o}-\frac{%
r^{2}\omega _{o}^{2}}{c^{2}}}}  \label{detauearth}
\end{equation}
For not too high observer's velocities and radii of the order of the size of
the Earth (\ref{detauearth}) becomes: 
\begin{equation}
\delta \tau \simeq \frac{4\pi }{c}r\left( \frac{r\omega _{o}}{c}+\frac{%
r^{3}\omega _{o}^{3}}{2c^{3}}+G\frac{M}{c^{3}}\omega _{o}-2G\frac{M}{%
c^{2}r^{2}}a\right)  \label{approssi}
\end{equation}

One easily recognizes, in the order, the traditional Sagnac term, then a
purely special relativistic correction, a ''Schwarzschild'' correction and
finally an angular momentum correction which is independent from the
velocity of the observer.

When the observer is freely orbiting on a circular trajectory the starting
point is formula (\ref{completo}); developing and keeping only the lowest
order terms up to the one containing the angular momentum, the result is 
\[
\delta \tau _{\pm }\simeq \frac{4\pi }{c^{2}}\left( \pm \sqrt{GMr}\pm \frac{1%
}{2c^{2}}\sqrt{\frac{G^{3}M^{3}}{r}}\pm G\frac{M}{c^{2}}\sqrt{\frac{GM}{r}}%
-2a\frac{GM}{cr}\right) 
\]
Now summing the time differences for the two rotation directions we obtain 
\begin{equation}
\Delta \left( \delta \tau \right) =\delta \tau _{+}+\delta \tau _{-}=-16\pi 
\frac{GM}{c^{3}r}a  \label{deltadelta}
\end{equation}

The order of magnitude of this quantity is $\sim 10^{-16}$ s.

\section{Conclusion}

We have given a general geometric treatment of the problem of the phase
shift of electromagnetic waves moving in a closed circular path about a
massive spinning body. Applying the general methodology to the terrestrial
environment reproduces approximate results obtainable by various alternative
techniques. In terms of time differences accumulated during round trips
around the Earth the numbers remain extremely small and partly smaller than
some values typical of the gclock effect. The idea of using electromagnetic
waves in some orbital Sagnac type experiment deserves however some attention
because of a few advantages light has.

To begin with, if a chain of orbiting mirrors or transponders are considered
it is possible to use them simultaneously for prograde and retrograde rays,
thus ensuring the coincidence of the two trajectories and the significance
of the time of flight difference.

Next we can notice that, even for an ''experimental apparatus'' as big as
the orbit of a satellite, the time of flight of light for a round trip is $%
\sim $ $0.1$ s. In that time the observer on a spacecraft could move for a
few hundred meters at most and an observer on the surface of the Earth would
move much less than that; this fact in principle allows for experiments at
different values of the equivalent angular velocity $\omega _{o}$, to be
maintained for a few seconds at most.

In fact in the time delay (\ref{approssi}) three terms depend on $\omega
_{o} $ and one is independent from it: in principle determining the values
for positive or negative $\omega _{o}$ close to zero could allow for the
detection of the invariant $a$ term. The scheme one can imagine is an
observer on the Earth moving westward at variable linear speeds of the order
of $\sim 10$ m/s and sending and receiving signals to and from a ring of
orbiting mirrors.

Finally, varying the wavelength of the waves could in principle allow to
tune the phase difference originated by the $a$ term, making it correspond
to a predefined number or fraction of interference fringes and evidencing
its presence with respect to the special relativistic or purely
gravitational shifts.

\end{document}